\UseRawInputEncoding
 \documentclass[conference]{IEEEtran}
\IEEEoverridecommandlockouts
\usepackage[left=1.62cm,right=1.62cm,top=1.9cm]{geometry}
\usepackage{cite}
\usepackage{amsmath,amssymb,amsfonts}
\usepackage{algorithmic}
\usepackage{graphicx}
\usepackage{textcomp}
\usepackage{xcolor}
\def\BibTeX{{\rm B\kern-.05em{\sc i\kern-.025em b}\kern-.08em
    T\kern-.1667em\lower.7ex\hbox{E}\kern-.125emX}}

\makeatletter
\newcommand{\linebreakand}{%
  \end{@IEEEauthorhalign}
  \hfill\mbox{}\par
  \mbox{}\hfill\begin{@IEEEauthorhalign}
}
\makeatother

\begin{document}

\title{CVD Growth of Tin Selenide Thin Films for Optoelectronic Applications\\
\thanks{*email: kbvinayak@iisertvm.ac.in}
}

\author{\IEEEauthorblockN{Shivam Tyagi}
\IEEEauthorblockA{\textit{School of Physics} \\
\textit{Indian Institute of Science Education and Research}\\
Thiruvananthapuram, India \\
}
\and
\IEEEauthorblockN{Shubham Yadav}
\IEEEauthorblockA{\textit{School of Physics} \\
\textit{Indian Institute of Science Education and Research}\\
Thiruvananthapuram, India \\
}
\linebreakand
\IEEEauthorblockN{Vaibhav Wani}
\IEEEauthorblockA{\textit{School of Physics} \\
\textit{Indian Institute of Science Education and Research}\\
Thiruvananthapuram, India \\
}
\and
\IEEEauthorblockN{ Shivam Singh}
\IEEEauthorblockA{\textit{School of Physics} \\
\textit{Indian Institute of Science Education and Research}\\
Thiruvananthapuram, India \\
}
\linebreakand
\IEEEauthorblockN{ Soumya Biswas}
\IEEEauthorblockA{\textit{School of Physics} \\
\textit{Indian Institute of Science Education and Research}\\
Thiruvananthapuram, India \\
}
\and
\IEEEauthorblockN{Krishna Nand Prajapati}
\IEEEauthorblockA{\textit{School of Physics} \\
\textit{Indian Institute of Science Education and Research}\\
Thiruvananthapuram, India \\
}
\linebreakand
\IEEEauthorblockN{ Vinayak B Kamble*}
\IEEEauthorblockA{\textit{School of Physics} \\
\textit{Indian Institute of Science Education and Research}\\
Thiruvananthapuram, India \\
kbvinayak@iisertvm.ac.in}
}

\maketitle

\begin{abstract}
Tin Selenide (SnSe) thin films were grown onto glass and alumina substrates by Chemical Vapor Deposition (CVD) method. The structural, micro-structural and morphological characterizations of the as grown thin films were investigated using XRD, SEM and Raman spectroscopy  which reveals that the films on glass are phase pure oriented SnSe while those on Alumina are polycrystalline SnSe with SnSe$_2$ impurity phase. The optical properties of the films grown onto glass substrate were studied by UV-vis spectroscopy. The optical band gap calculated is 1 to 1.3 eV for indirect and direct  transition in film deposited on glass substrates. The Arrhenius plots of the two films show very different thermal activations i.e. 0.088 eV for tin vacancy acceptor level close to valance band maxima in pure SnSe and 0.44 eV for mid gap selenium vacancies of SnSe$_2$. Photoresponse was observed by illuminating the sample (Glass and alumina Substrate) using white and UV light (400 nm) for a fixed time pulses. The deposited onto Alumina substrate were found to show better photoresponse due to SnSe/SnSe$_2$ p-n heterojunctions.
\end{abstract}


\section{Introduction}
Two-dimensional (2D) layered chalcogenide materials (LCMs), especially the IVA-VIA group, have attracted tremendous attention to the demands for next generation electronics and optoelectronics. Bulk SnSe is a p-type semiconductor with an indirect band gap of $\sim$ 0.9 eV and a direct band gap of $\sim$ 1.3 eV \cite{kumar2021tin, zhang2015plasma}.  SnSe is a versatile material with a wide variety of applications in optoelectronics, sensors, thermoelectrics etc\cite{kumar2021tin}. Due to its small bandgap and layered structure it is exploited for a broadband photodetector application as it absorbs most of the NIR, visible and UV spectrum \cite{kang2018photodetector}.

\begin{figure}[]
\centerline{\includegraphics[width=8cm]{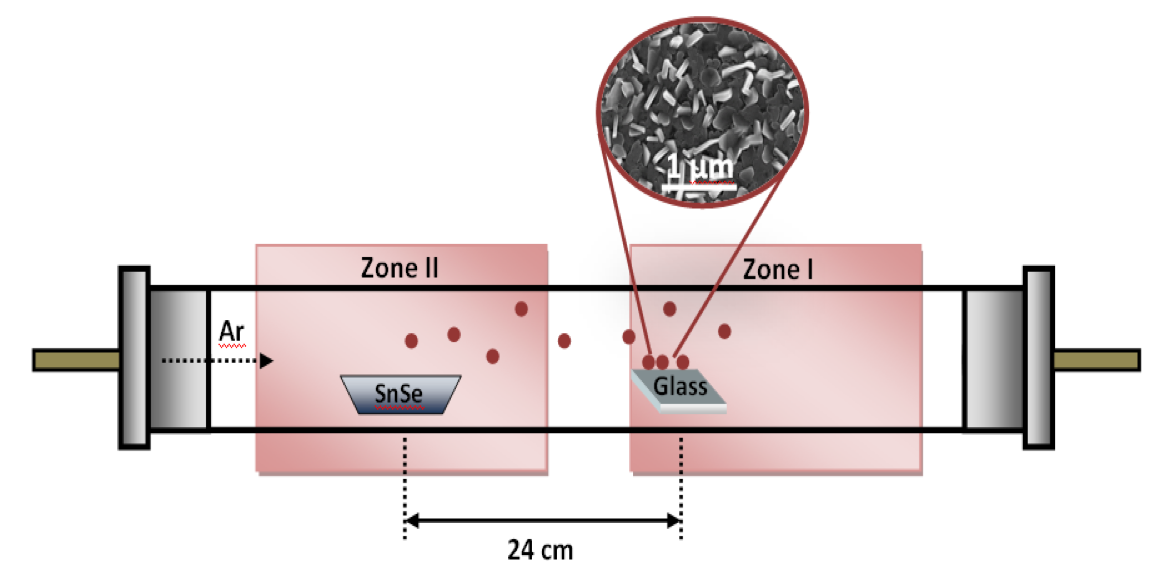}}
\caption{Systematic illustration of SnSe thin film deposition by Chemical deposition method.}
\label{fig1}
\end{figure}

Recently, several research groups have reported that pure phase 2D SnSe and SnSe$_2$ can be synthesized by methods such as micromechanical exfoliation, solution based methods, vapor deposition, plasma-assisted method etc \cite{gong2018extremely, davitt2020crystallographically, zhang2015plasma}. Nevertheless, the challenge lies in controlling the phase purity to get single-phase SnSe or SnSe$_2$ because $Sn^{2+}$ in the precursor is readily oxidized into $Sn^{4+}$, resulting in a mixture of SnSe and  $SnSe_2$.
As thin film technology can be integrated into most emerging electronic and optoelectronic devices, the development of thin film phase selection is of great significance. Here, our aim is to go for the deposition of single layer pure phase SnSe by CVD process. In this report, we are preparing SnSe nanopowder as a precursor for SnSe thin film deposition by using a simple hydrothermal route mentioned below. The prepared SnSe precursor is used directly in CVD growth SnSe.

\begin{figure}[]
\includegraphics[width=9cm]{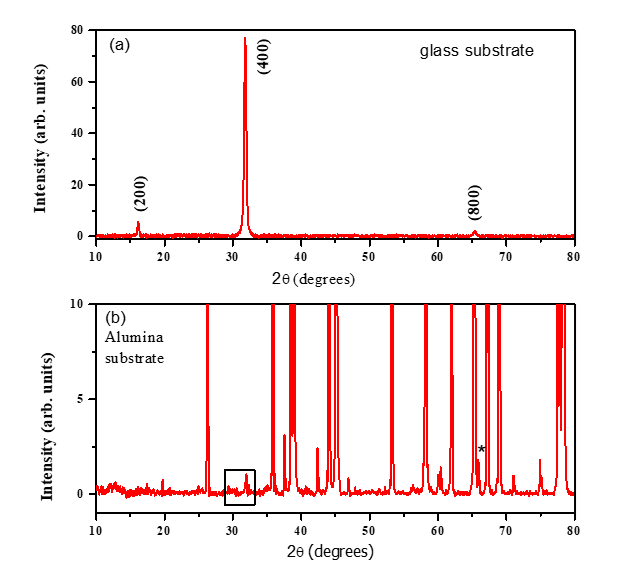}
\caption{The X-ray diffraction patterns of SnSe thin films deposited on (a) glass substrate and (b) Alumina substrate. (The SnSe peak has been shown with square in (b)}
\label{xrd}
\end{figure}

He et al.\cite{huang2015designing} used Se and SnSe powder as precursors to prepare SnSe and  $SnSe_2$ nanosheets by the method of vapor deposition (VD). The phase transition of a similar system of tin sulfides was realized by the addition of $H_2$ \cite{ahn2015deterministic} and regulating the temperature \cite{mutlu2016phase} in a CVD process. Boscher et al. \cite{boscher2008atmospheric} obtained SnSe and $SnSe_2$ thin films on glass via atmospheric pressure (AP) CVD. Huang et al. \cite{huang2015designing} also studied the effect of growth temperature on the phase of the products. From the above results \cite{boscher2008atmospheric, huang2015designing} it is evident that the temperature plays a crucial role in the phase formation of tin selenide films. However, the critical temperature of the tin selenide phase transition varied from 650$^o$C \cite{boscher2008atmospheric} to 510$^o$C \cite{huang2015designing}. In a process where VD preparation was carried out in the atmosphere, the phase transition temperature raised to 650$^o$C \cite{boscher2008atmospheric}. For the top-down preparation via VD,  SnSe$_2$ films were beneficial to be prepared at reaction temperature between 300$^o$C and 470$^o$C, and SnSe films could be grown at 230$^o$C to 280$^o$C or 500$^o$C to 570$^o$C \cite{fernandes2013thermodynamic}.

In this paper, we report the fabrication of phase pure SnSe thin films through optimization of CVD process parameters. The films deposited on amorphous (glass) and crystalline (alumina) substrates show very different crystallinity and hence markedly different transport properties. Although the films deposited on glass are highly textured along (400) i.e. the crystalline axis of 2D layers of the structure, a better photoresponse is observed for polycrystalline films on Alumina substrates.

\begin{figure}[]
\includegraphics[width=9cm]{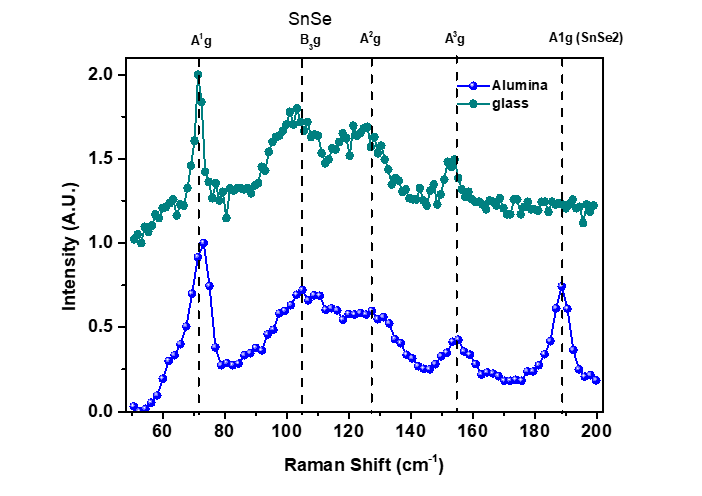}
\caption{Raman Spectra of the films deposited on (a) glass substrate and (b) Alumina substrate.}
\label{raman}
\end{figure}

\begin{figure*}[]
\center
\includegraphics[width=14cm]{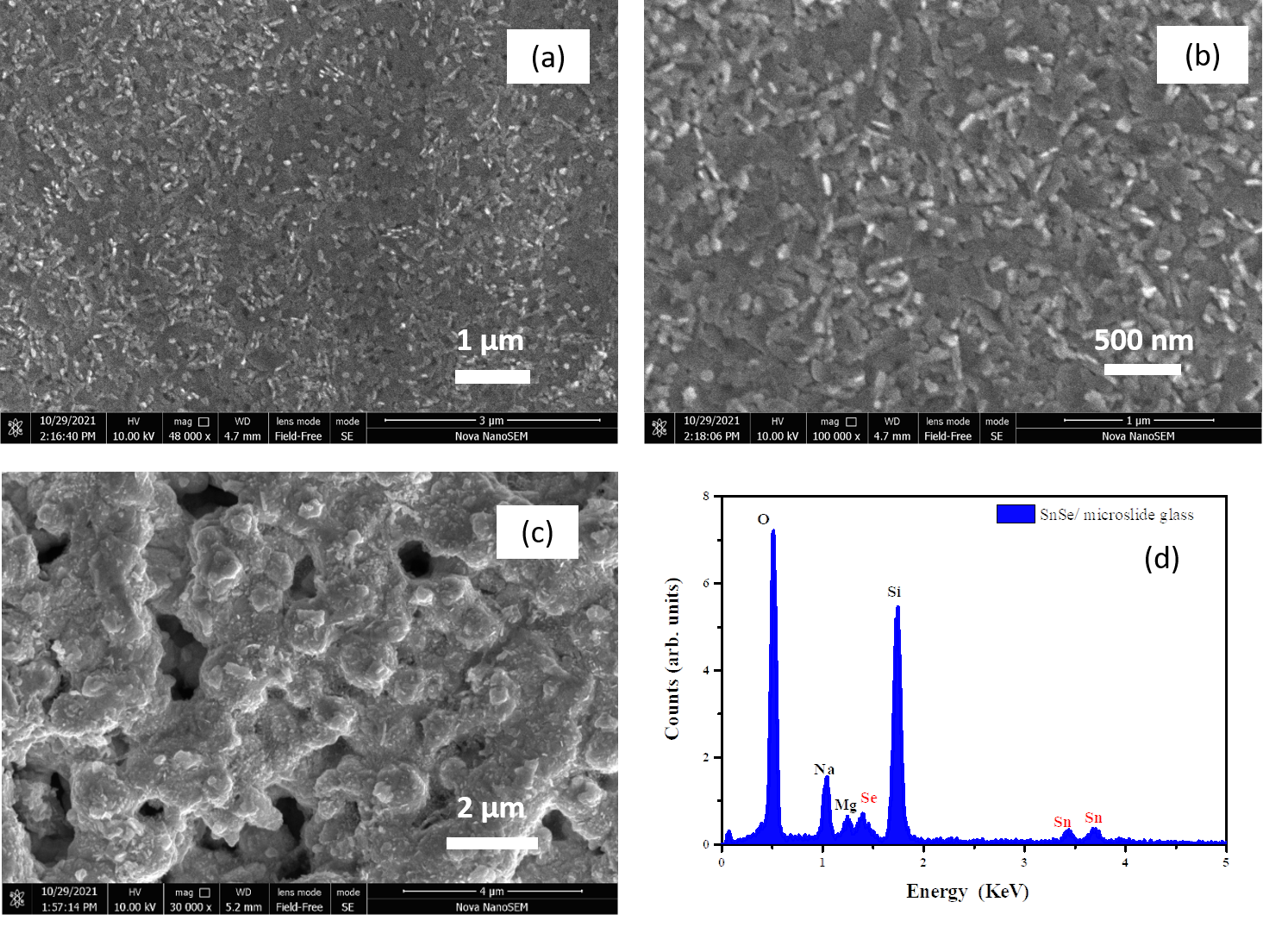}
\caption{The SEM micrographs of SnSe thin films deposited on (a and b) on glass substrate and (c) on Alumina substrate. (d) The EDS spectrum of SnSe thin film  showing only Sn and Se along with composition of glass}
\label{SEM}
\end{figure*}

\section{Experimental}
Tin (II) chloride dihydrate ($SnCl_2.2H_2O$, 98\%, MERCK), selenium powder (Se, 99.99\%, Sigma-Aldrich), Potassium Hydroxide pellets (KOH, 84\%, MERCK) and sodium hydroxide (NaOH, 98\%, Sigma Aldrich) were used for the synthesis without any further purification.
SnSe nanoplates were successfully synthesized using simple hydrothermal method route \cite{nerella2021effect}. First a mixture of $SnCl_2.2H_2O$ (1 mmol, 0.2253 gm) was dissolved in deionized water (40 ml). The solution was then stirred at room temperature for 5 min, followed by the addition of KOH (10 mmol, 0.5611 gm) and stirred for another 15 min unless a transparent solution was obtained. Se powder (1mmol, 0.0789 gm) was added to the above solution and stirred for 15 min at room temperature and the solution turned grayish black. Furthermore, NaBH4 (5 mmol, 0.1891 gm) was mixed in the solution and stirred for another 30 min; this turned the solution to pale yellow. The mixture was then transferred to a teflon-lined stainless-steel autoclave of 50 ml capacity and sealed tightly and heated to 170 $^o$C for 12h. After cooling to room temperature, the black precipitate was collected and thoroughly washed with ethanol and deionized water for several times. Finally, the product was vacuum dried for 10 h and collected as powder. 

As prepared SnSe (nanoplates) powder was used as a precursor for SnSe growth by Chemical vapour deposition method as shown in FIG \ref{fig1}. The glass and silicon substrates were cleaned in soap water, distilled water, acetone and IPA. The Experiment was performed in a deposition chamber that has two zone furnaces: zone I and zone II. To grow SnSe thin film, precursor was kept in an alumina boat in the centre of Zone II. The cleaned substrates were placed in Zone I at (21 cm - 28 cm) distance from the precursor in Zone I (FIG \ref{fig1}). Further, the tube was pumped down to remove the extra oxygen present in the deposition chamber. The temperature gradient was maintained at 800$^o$C for Zone II and 350$^o$C for Zone I \cite{lu2021phase}. 
To grow SnSe thin films Ar flow was maintained at 100 SCCM by mass flow controller and pressure was controlled at 1 mbar in the deposition chamber. After 15-20 minute of growth time, the furnace was left to cool down naturally to room temperature. Here SnSe thin films was deposited on glass and alumina substrates.

\section{Results}
\subsection{structural studies}
X-Ray diffraction patterns were obtained and analyzed to find the phase structure of SnSe thin films (FIG \ref{xrd}). These corresponding peaks are consistent with peaks of orthorhombic SnSe reported in the literature \cite{8605911}.  The XRD pattern shows sharp peaks at (200), (400) and (800) reflections indicating a (h00) preferred orientation in film deposited on glass. This confirms pure phase SnSe thin films are at 24 cm distance from the precursor.  Whereas, the XRD patterns of films deposited at closer and further distances show mixed impurity of SnSe$_2$ \cite{ramasamy2015phase}. On the other hand, the films deposited on alumina substrate (FIG \ref{xrd}(b)) did not show any such preferred orientation. It is possible that due to the small thickness of films compared to the crystalline substrate most reflections are observed from the alumina substrate. Nevertheless, it certainly rules out any possibility of texture (preferred orientation) in the films. The rhombohedral lattice of Alumina has no crystalline relationship with that of orthorhombic SnSe and hence no strong orientation observed on the polycrystalline substrate. The position of most probable peaks of SnSe are marked with the box in FIG \ref{xrd}(b). Whereas the only new peak observed in the film compared to Alumina has been marked by an asterisk (*), which matches (800) peak position.

Raman Spectroscopy measurement was performed to investigate the quality of as-grown SnSe thin films. Five distinct peaks were observed at 70.91, 104.39, 120.77, 153.01, 183.28 $cm^{-⁻1}$ (FIG \ref{raman}). Here, $A_g$ and $B_{3g}$ are two rigid shear modes of a layer with respect to its neighbour and confirm the planar vibration modes of orthorhombic phase SnSe.[\cite{8605911, xu2017plane}  The highest intensity peak was observed at 70.91$cm^{-⁻1}$ which corresponds to $A_{1g}$  vibrational mode.\cite{xu2017plane}. The last peak ($A_{1g}$) is attributed  to the vibrational mode of SnSe$_2$. Thus it may be confirmed that both glass as well as Alumina substrate films have SnSe formed while the alumina substrate film has SnSe$_2$ impurity phase.

\subsection{Film morphology}
The Scanning Electron Microscopy  images of SnSe thin films on glass as well as alumina substrates are shown in FIG \ref{SEM}.It may be seen from the figure that films on the glass have smoother flat platelets compared to the rough surface of alumina substrate films. The roughness could be due to inherent roughness of the alumina substrate and a conformal coating deposited by the CVD method. These results are in good agreement with results obtained in XRD \cite{lu2021phase}. The typical film thicknesses measured for glass substrates using a profilometer are 80-100 nm.

The Energy Dispersive Spectroscopy (EDS) spectra of the films on glass is shown in FIG \ref{SEM}(d). It may be seen that along with the constituent of soda lime glass, only Sn and Se peaks are observed. The typical relative percentages of Sn and Se obtained are summarized in Table below.

\begin{table}[b]
\center
\caption{The summary of Elemental quantification using Energy dispersive spectroscopy for SnSe films on glass and alumina}
\begin{tabular}{|c|c|c|c|}
\hline
element 	& Glass substrate & Alumina substrate \\
\hline
Sn (L) & 1.53 & 0.35 \\
\hline
Se (K) & 2.58 & 0.17 \\
\hline
Sn:Se & 0.6:1 & 2.05:1 \\
\hline

\end{tabular}
\label{table_EDS}
\end{table}

From the EDS spectra quantification, it is apparent that film deposited on glass has lower Sn content while that of alumina substrate films has it in excess. It may be noted that these numbers cannot be taken sacrosanct and are just indicative of relative composition. A complete mapping of EDS or XPS could provide more accurate quantifications.

\subsection{Optical studies}
\begin{figure}[t]
\center
\includegraphics[width=9cm]{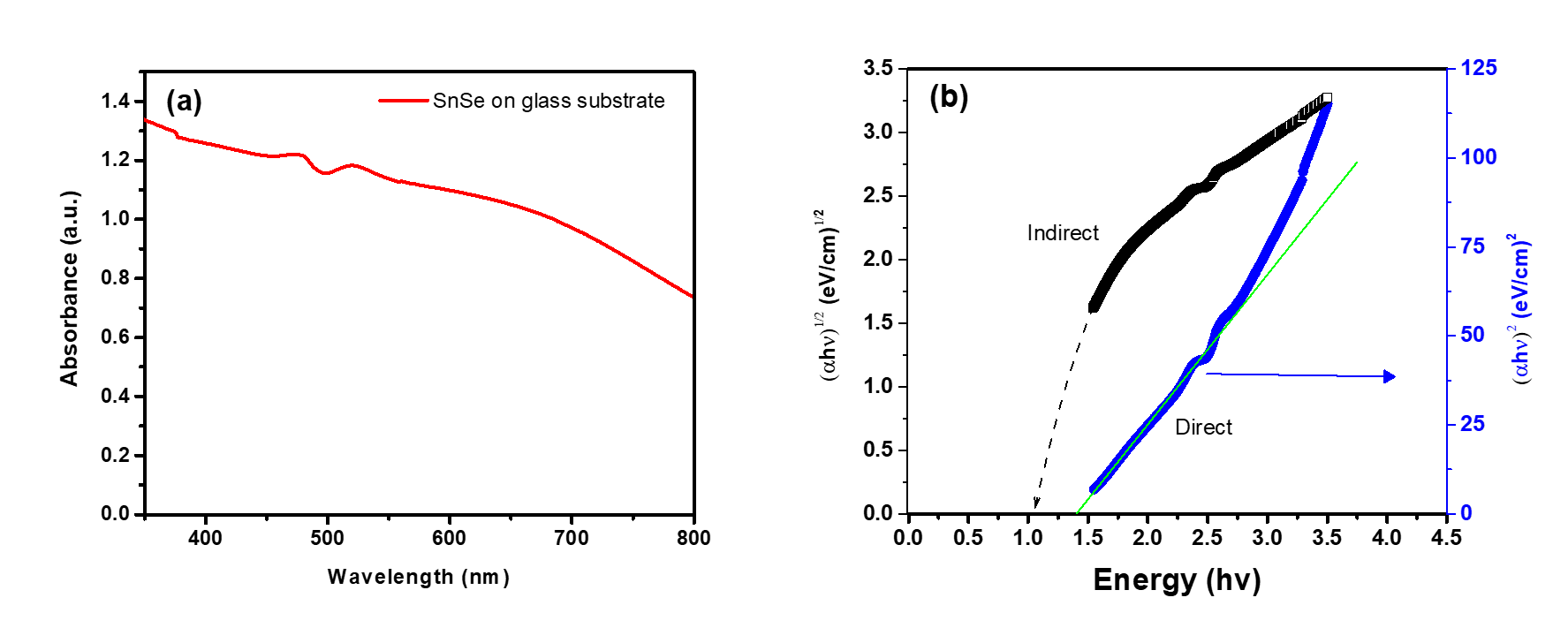}
\caption{(a) The UV-vis absorption spectra of SnSe thin films deposited on glass substrate and (b) its Tauc plot showing band gap vaue of about 1.4 eV. }
\label{abs}
\end{figure}
The optical absorption was studied by UV-Visible absorbance spectrum ranging from 350 - 800 nm. The absorbance spectrum is shown in FIG \ref{abs}(a). The films show high absorbance over the entire uv-visible spectrum. The value of the energy bandgap of SnSe thin films deposited Glass Substrate was calculated from the Tauc plot using the following relation:  
\begin{equation}
\alpha h \nu = A(h \nu - E_g )^{1/n} 
\end{equation}
                                                                   
where, $\alpha$ is the absorption coefficient, A is the proportionality constant, h$\nu$ is incident photon energy and $E_g$ is the band gap of the material. For direct band gap, value of n is taken to be ½ and for indirect n is 2. So, $E_g$ was determined via extrapolating the linear region of the curve of ($α\alpha h \nu)^{1/n}$ versus $h\nu$ cuts the abscissa. For the indirect transition, the range of data measured was insufficient and hence extrapolated as shown. In FIG \ref{abs}(b) it may be noticed that the value of band gaps for SnSe thin film deposited is found to be about $\sim$1 eV and 1.3 eV for indirect and direct transitions respectively. These are good agreements as reported in the literature\cite{kumar2021tin}.

\subsection{Transport measurements}
\begin{figure}[b]
\center
\includegraphics[width=8cm]{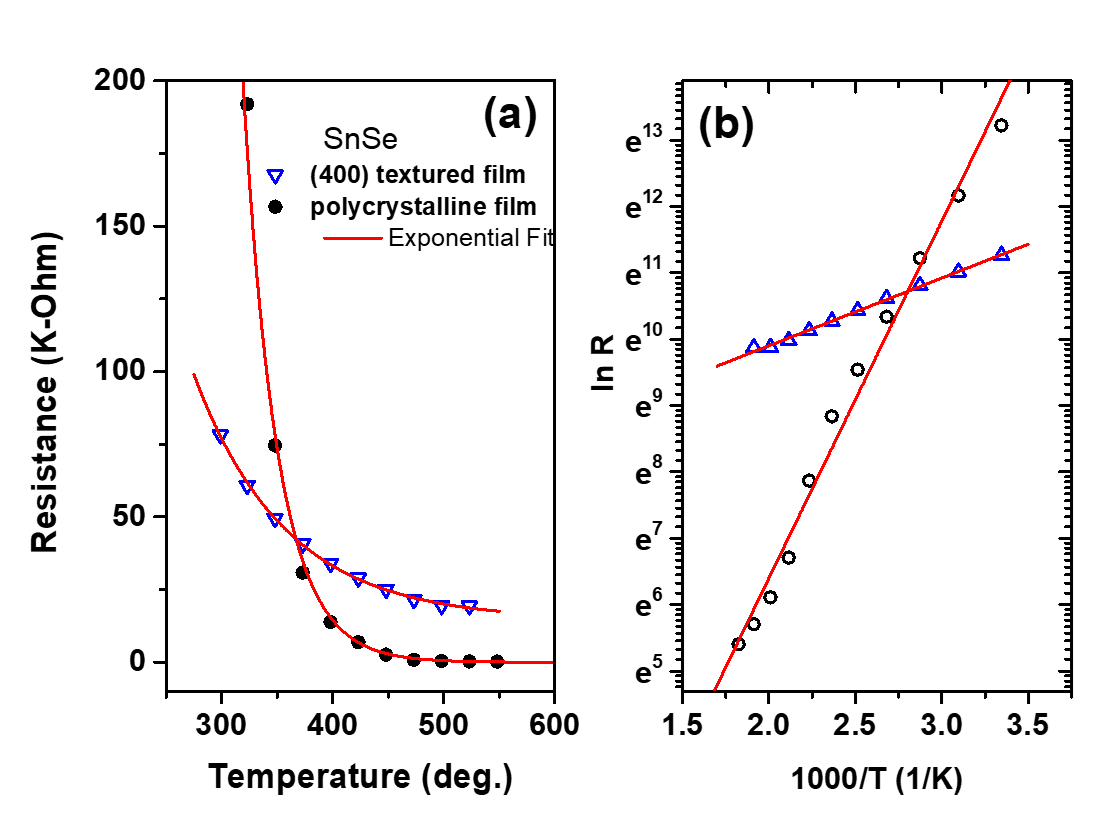}
\caption{(a) The variation of resistance with temperature for film on glass ((400) textured) and Alumina (polycrystalline. (b) The Arrhenius plot of the two films showing very different thermal activation energies of 0.46 eV and 0.088 eV for glass and alumina substrates respectively.}
\label{R-T}
\end{figure}

CVD grown SnSe thin films conduction mechanism was identified using temperature dependent current voltage characteristics. I-V characteristics were taken for glass and alumina substrate samples. SnSe is reported to show anomalous semiconductor to metal like and again semiconducting trend in resistivity due to bipolar effect and structural transition at 550 K and 800 K respectively \cite{loa2018critical}. However, within the range of temperature studied in this work, both the samples showed semiconducting nature as seen in FIG \ref{R-T}(a). Nevertheless, the room temperature resistance of textured film was very low compared to that of polycrystalline film. This could be due to lattice mismatch at the grain boundary in polycrystalline film \cite{zhao2014ultralow, burton2018thin}. Whereas the textured film shows (400) orientation which depicts the 2D nature of films (as SnSe has a layered structure along the a-axis), the lattice mismatch is minimal. 
On the other hand, when temperature is raised, the carriers are excited thermally and the thermal excitation energy is calculated using the Arrhenius relation,
\begin{equation}
R = R_0\; exp(E/k_BT)
\end{equation}
where, $R_0$ is resistance at reference temperature (T=0) and $k_B$ is Boltzmann constant. E is the energy of thermal activation.

SnSe shows p-type conductivity due to presence of tin vacancies ($V_{Sn}^{2-}$) which gives rise to an acceptor level just above the valence band \cite{wei2018achieving}. The activation energy calculated for glass substrate is found to be 88 meV which could be due to activation carriers to defect level state within the gap. Due to excess defects ($V_{Sn}$) the film shows very high conductivity and lower activation energy. Whereas the activation energy obtained for polycrystalline films having SnSe$_2$ impurity is 0.44 eV which is almost half the bandgap value and hence depicts a thermal activation across the band edge. However, in that case thermal activation should result in bipolar conductivity which could be responsible for a steep fall of resistance with temperature due to bipolar conduction. Such activation is observed in As doped SnSe wherein substitution occurs at Sn site and the dominant defect is Se vacancies ($V_{Se}^{2+}$) which lies approximately 0.4 eV above the valence band edge. \cite{Navratil2018} SnSe$_2$ being an n-type conductor, the presence of impurity phases results in heterojunction barriers. 

\subsection{Photoresponse}
\begin{figure}[]
\center
\includegraphics[width=9cm]{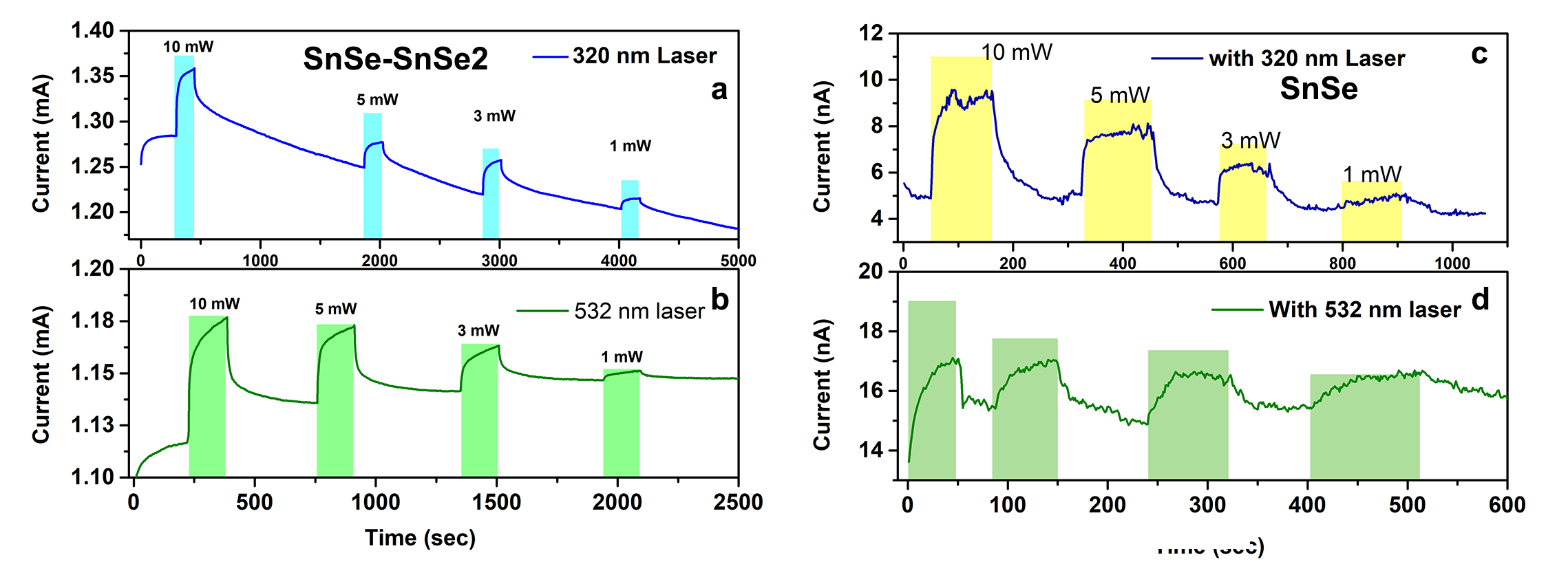}
\caption{The phtoresponse  for polycrystalline films on alumina substrate (a) 320 nm and (b) 532 nm and phtoresponse  for oriented films on glass substrate (c) 320 nm and (d) 532 nm.}
\label{photores_times}
\end{figure}
\begin{figure}[b]
\center
\includegraphics[width=9cm]{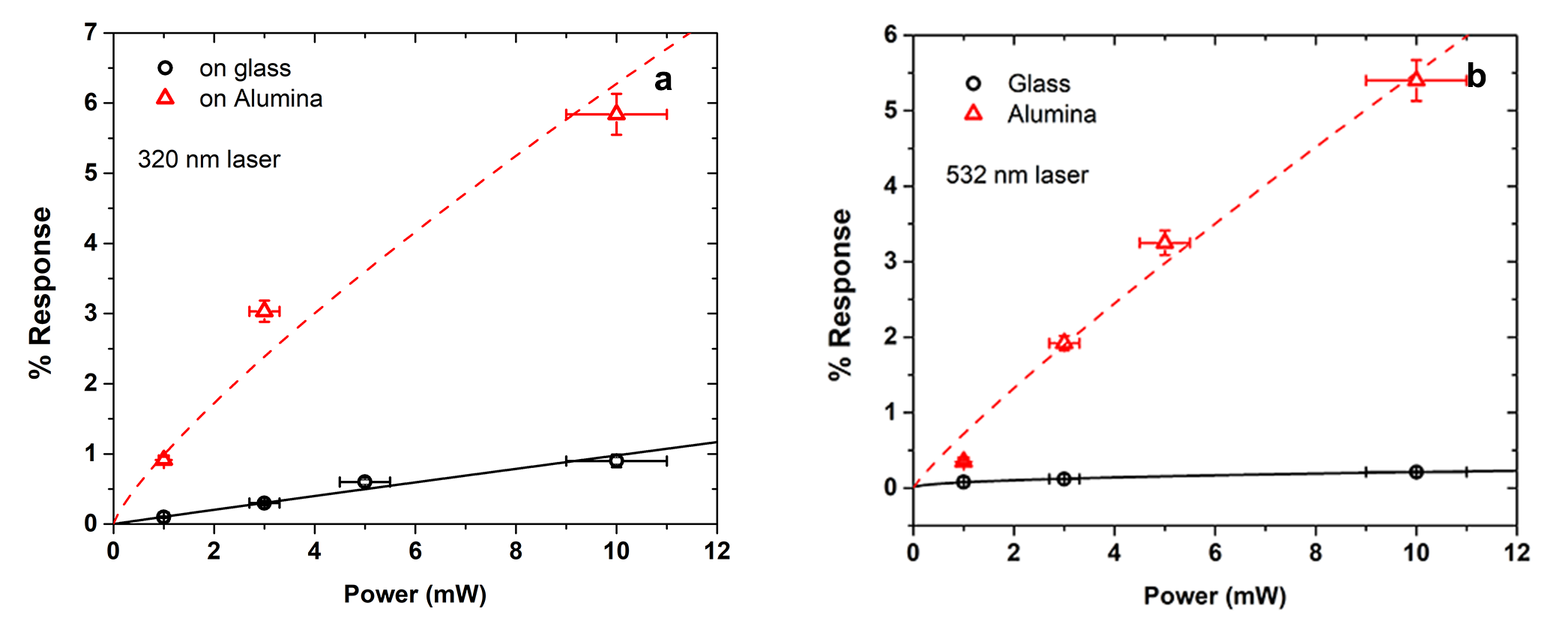}
\caption{The phtoresponse to (a) UV light 320 nm and (b) green light 532 nm for textured films deposited on Glass substrate and polycrystalline film on alumina substrate.}
\label{photores_Gl}
\end{figure}

Here we observed the dark and photocurrent at a fixed voltage under the illuminating UV light (320 nm) and green (532 nm) for SnSe deposition alumina substrate [FIG \ref{photores_times}(a) and (b)] as well as on glass substrate [Fig \ref{photores_times}]. There was a sharp photocurrent increase with light intensity in both the cases, however, the overall time taken to saturation (response) and recovery was lower for that of glass substrate film. When normalized with baseline current, the response  i.e. $\Delta I/I$ for glass substrate film turned out to be low compared to alumina substrate film. (see table \ref{photores_Gl}). 

This difference in response could be due to the effect of p-n junction hetero-interface which formed due to the impurity phase of SnSe$_2$ (n-type) in the matrix of SnSe (p-type) \cite{gowthamaraju2021augmentation, zhong2020}. Nevertheless, the response time due to this showed two time constant like one initial fast rise and then a slow rise going to saturation. For UV light photo current response was larger than white light.
 
\section{Conclusion}
Tin selenide (SnSe) powder was prepared by using a hydrothermal method. This was used as a precursor for thin film deposition by CVD method. Here, we deposited SnSe on glass and Alumina substrates. It was found that different type of substrates influence on SnSe thin films crystallinity and phase purity. XRD, Raman and SEM confirm the orthorhombic structure of preferred orientation with layered structure of the SnSe thin films on glass. On the other hand Alumina substrate films show no preferred orientation and impurity phase $SnSe_2$ phase. The optical band gap varies nearly 1 eV to 1.17 eV. The R-T measurements show that the glass substrate films have very low thermal activation energy with high conductivity, while that of alumina substrate films show almost half the bandgap energy as activation energy and possible bipolar effect. Photoresponse was observed in SnSe thin deposition onto alumina as well as glass substrate using white light and UV light. Here SnSe deposited onto Alumina substrate showed  better photosensing properties due to SnSe/SnSe$_2$ heterojunction inhomogeneities. Thus, a facile synthesis method is reported for fabrication of SnSe thin films and potential photosensing studies have been explored. Further studies quantifying the photoresponse are ongoing.

\section*{Acknowledgment}

The authors are thankful to Science and Engineering Research Board, Govt. of India for the funding (EEQ/2018/000769).

\bibliographystyle{IEEEtran}
\bibliography{mybib}

\end{document}